\documentclass[aps,prb,reprint,floatfix,superscriptaddress]{revtex4-2}

\usepackage{amsmath,amsfonts,amssymb} 
\usepackage{graphicx} 
\usepackage[dvipsnames,table]{xcolor} 
\usepackage{braket} 

\usepackage{algorithm}
\usepackage{algpseudocode}

\usepackage[breaklinks, 
colorlinks, linkcolor=Blue, citecolor=Blue, urlcolor=Blue]{hyperref} 

\usepackage{enumitem}

\usepackage{color}
\usepackage{makecell}
\usepackage{booktabs}
\usepackage{dashrule}

\def\be{\begin{equation}} 
\def\ee{\end{equation}}

\newcommand{\hn}{\hat n}

\newcommand{\HSS}{\hat S^2}
\newcommand{\CR}[1]{\hat a^{\dagger}_{#1}}
\newcommand{\AN}[1]{\hat a_{#1}}

\newcommand{\EE}[2]{\hat E_{#2}^{#1}}
\newcommand{\XX}[2]{\hat X_{#2}^{#1}}

\newcommand{\bea}{\begin{eqnarray}}
\newcommand{\eea}{\end{eqnarray}}

\newcommand{\HNe}{\hat N_e}
\newcommand{\HSz}{\hat S_z}



\begin{document}

\title{Planted Solutions in Quantum Chemistry: Generating Non-Trivial Hamiltonians with Known Ground States}

\author{Linjun Wang}
\affiliation{Department of Physical and Environmental Sciences,
University of Toronto Scarborough, Toronto, Ontario, M1C 1A4,
Canada}
\affiliation{Chemical Physics Theory Group, Department of Chemistry,
University of Toronto, Toronto, Ontario, M5S 3H6, Canada}

\author{Joshua T. Cantin}
\affiliation{Department of Physical and Environmental Sciences,
University of Toronto Scarborough, Toronto, Ontario, M1C 1A4,
Canada}
\affiliation{Chemical Physics Theory Group, Department of Chemistry,
University of Toronto, Toronto, Ontario, M5S 3H6, Canada}

\author{Smik Patel}
\affiliation{Department of Physical and Environmental Sciences,
University of Toronto Scarborough, Toronto, Ontario, M1C 1A4,
Canada}
\affiliation{Chemical Physics Theory Group, Department of Chemistry,
University of Toronto, Toronto, Ontario, M5S 3H6, Canada}

\author{Ignacio Loaiza}
\affiliation{Department of Physical and Environmental Sciences,
University of Toronto Scarborough, Toronto, Ontario, M1C 1A4,
Canada}
\affiliation{Chemical Physics Theory Group, Department of Chemistry,
University of Toronto, Toronto, Ontario, M5S 3H6, Canada}

\author{Rick Huang}
\affiliation{Department of Physics,
University of Toronto, Toronto, Ontario, M5S 3H6, Canada}  

\author{Artur F. Izmaylov}
\email{artur.izmaylov@utoronto.ca}
\affiliation{Department of Physical and Environmental Sciences,
University of Toronto Scarborough, Toronto, Ontario, M1C 1A4,
Canada}
\affiliation{Chemical Physics Theory Group, Department of Chemistry,
University of Toronto, Toronto, Ontario, M5S 3H6, Canada}

\date{\today}

\begin{abstract}
\section*{Abstract}
Generating large, non-trivial quantum chemistry test problems with known ground-state solutions remains a core challenge for benchmarking electronic structure methods. Inspired by planted-solution techniques from combinatorial optimization, we introduce four classes of Hamiltonians with embedded, retrievable ground states. These Hamiltonians mimic realistic electronic structure problems, support adjustable complexity, and are derived from reference systems. Crucially, their ground-state energies can be computed exactly, provided the construction parameters are known. To obscure this structure and control perceived complexity, we introduce techniques such as killer operators, balance operators, and random orbital rotations. We showcase this framework using examples based on homogeneous catalysts of industrial relevance and validate tunable difficulty through density matrix renormalization group (DMRG) convergence behavior. Beyond enabling scalable, ground-truth benchmark generation, our approach offers a controlled setting to explore the limitations of electronic structure methods and investigate how Hamiltonian structure influences ground state solution difficulty.

\end{abstract}

\maketitle

\section{Introduction}

Quantum chemistry has made extensive use of benchmark data sets for several decades \cite{fisher_diagnostics_2016,nagy_state_2024} to assess various methods, from density functional theory (DFT) \cite{curtiss_assessment_2005,sousa_general_2007,karton_highly_2008,Mardirossian_thirty_2017,karton_highly_2025} and time-dependent density functional theory \cite{laurent_tddft_2013}, to quantum algorithms \cite{mccaskey_quantum_2019}, coupled cluster (CC) methods \cite{schreiber_benchmarks_2008},  and beyond \cite{nagy_state_2024,guner_standard_2003,zhao_benchmark_2008,zhao_benchmark_2009,haunschild_new_2012,karton_w417_2017}. These benchmarks are used to test the accuracy of the methods and to fit parametrized methods \cite{sousa_general_2007,karton_highly_2008,karton_good_2025}. Benchmarks help researchers identify where methods work well and where they do not \cite{schreiber_benchmarks_2008,nagy_state_2024,obuchi_sparse_2016} and they help drive method development by defining performance standards \cite{qb_benchmark_repo,gsee_benchmark_paper}.

Accurate reference energies are critical for benchmark data sets to be reliable \cite{karton_good_2025,savin_acknowledging_2020}. However, in many cases the well-known reference energies 
correspond to small or non-industrially relevant molecules. Advanced methods, such as the W4 method \cite{karton_highly_2008,karton_w417_2017}, semistochastic Heat-bath configuration interaction \cite{li_fast_2018}, or the density matrix renormalization group (DMRG) \cite{whiteDmrg1992,whiteDmrg1993, zhai2023block2, larsson_chromium_2022,wouters_density_2014,
bellonziFeasibilityAcceleratingHomogeneous2024}, can be used to produce excellent reference data, but these are computationally expensive. While relatively more accurate methods, such as CC, can be used to produce reference data to benchmark less accurate but computationally faster methods \cite{laurent_tddft_2013,fisher_diagnostics_2016}, such as DFT, this does not resolve the challenge of obtaining accurate 
benchmarks for the advanced methods themselves. Moreover, there is a need for  \textit{large} benchmark data sets, such as those needed to train machine learning models \cite{stein_advancing_2022}; this size requirement further increases the computational burden.

A similar reference data problem arises in the field of optimization. One strategy used in the field for decades to address this problem is the use of planted solutions \cite{bach_how_1983,pilcher_partial_1992,krzakala_hiding_2009,barthel_hiding_2002,angelini_limits_2023}. Planted solutions are artificially generated problems that have an easy-to-calculate reference value if the underlying structure or construction procedure details are known, but are typically difficult to solve without that knowledge. Prime factorization provides a conceptually simple example \cite{bach_how_1983}: choose two large  prime integers at random and then multiply them to obtain the composite integer to be factorized. This produces a hard problem to solve, factoring a large integer, but for which we already know the answer. 

Planted solutions have been used extensively for various optimization problems, from constraint satisfaction problems \cite{barthel_hiding_2002,krzakala_hiding_2009}, Ising and other binary optimization  \cite{kowalsky_3regular_2022,hen_equation_2019,hen_probing_2015,marshall_practical_2016,king_quantum_2019,gangat_hyperoptimized_2024,perera_chook_2021,hamze_near_2018,king_performance_2015,albash_demonstration_2018,hamze_wishart_2020,perera_computational_2020}, graph coloring \cite{angelini_limits_2023}, and compressed sensing problems \cite{obuchi_sparse_2016}. Planted solutions have also been developed for the ground state and dynamics of systems probed by nuclear magnetic resonance \cite{hen_darpa_talk_2025}, and the ground state problem of the Fermi-Hubbard model \cite{kojima2025FermiHubbard}. Several of these various planted solution types also have a difficulty level that can be tuned \cite{hen_probing_2015,king_quantum_2019,hamze_wishart_2020,hamze_near_2018,perera_computational_2020,angelini_limits_2023}, such as by changing the ruggedness of the optimization landscape \cite{king_quantum_2019}, frustration \cite{king_quantum_2019}, or overall complexity \cite{hamze_near_2018}; they can even display hardness ``phase'' transitions \cite{perera_computational_2020,angelini_limits_2023}.

One key recent use of planted solutions is to benchmark the performance of early-stage quantum computers ranging from quantum annealers \cite{kowalsky_3regular_2022,hen_probing_2015,marshall_practical_2016,king_performance_2015,albash_demonstration_2018} to NISQ devices \cite{bharti_noisy_2022,yeteraydeniz_benchmarking_2021}. In the case of fault tolerant algorithms, planted solutions have been used to predict the resources required to solve benchmark problems \cite{qb_benchmark_repo,gsee_benchmark_paper}.

Here, we propose several classes of tunable planted solutions for use in quantum chemistry, particularly in electronic ground-state energy estimation (GSEE) benchmark data sets. These planted solutions are pseudo-chemical Hamiltonians that respect the fundamental physical symmetries of a chemical system and are composed of chemically- and physically-motivated Hamiltonian structures. The structures chosen allow us to obtain the
ground state energies for system sizes that are of industrial
relevance. To avoid easy exploitation of these structures, we irreversibly obscure the Hamiltonian structures by adding killer operators \cite{loaiza_reducing_2023,loaiza_block_2023} and by conjugating the Hamiltonian with orbital rotations. Neither these killer operators nor the orbital rotations affect the ground-state energy of the Hamiltonian, providing obfuscation without loss of reference energy accuracy. The various classes of planted solutions and their flexible parametrizations further allow tuning the difficulty of the GSEE for a given planted 
solution. 

The rest of the paper is organized as follows. 
We describe the general features of these planted solutions and define five classes in Sec.~\ref{sec:planted_sols_description}, with a summary of the classes in Table~\ref{tab_planted_solution_summary}.
In Sec.~\ref{sec:results}, we demonstrate the use of several planted solutions, addressing GSEE with DMRG calculations. We also demonstrate that the difficulty of GSEE via DMRG for these planted solutions is readily tunable. 
We then conclude in Sec.~\ref{sec:conclusion}.

\section{Exactly solvable Hamiltonians}
\label{sec:planted_sols_description}

In this section, we will discuss some general features of the planted solutions as well as describe several types of exactly solvable Hamiltonians and how they can be used to generate planted solutions.

\subsection{General features}
\label{sec:general_features}

Let us take a Hamiltonian $\hat H$ with the structure $U^\dagger \hat H_\mathrm{sol} U$, where $\hat H_\mathrm{sol}$ represents a Hamiltonian with an easily obtainable ground-state energy (\textbf{sol}vable) and 
$U$ are orbital rotations that hide the $\hat H_\mathrm{sol}$ structure. $\hat H_\mathrm{sol}$ and $U$ are chosen such that $\hat H$ has the form of an electronic Hamiltonian:
\begin{align}
    \hat H &= U^\dagger \hat H_\mathrm{sol} U \\
    &= \sum_{p,q =1 }^N h_{pq}\EE{p}{q} + \sum_{p,q,r,s = 1}^N g_{pqrs}\EE{p}{q}\EE{r}{s} 
\end{align}
where $p$, $q$, $r$, and $s$ are indices for the $N$ \textit{spatial} orbitals; $h_{pq}$ are the one-electron integrals; $g_{pqrs}$ are the two-electron integrals; $\EE{p}{q} = \CR{p\alpha}\AN{q\alpha} + \CR{p\beta}\AN{q\beta}$ are the singlet excitation operators with $\CR{p\sigma}$ ($\AN{p\sigma}$) electron creation (annihilation) operators for orbital $p$ and spin z-axis projection $\sigma=\{\alpha,\beta\}$.

The orbital rotations $U$ take the form
\begin{align}
U &= \prod_{p>q}^N e^{\theta_{pq}(\EE{p}{q} - \EE{q}{p})} \label{eq:singlet_rotations}\\
 &= \prod_{P>Q}^{2N} e^{\theta'_{PQ}(\XX{P}{Q}\ - \XX{Q}{P})},
\end{align}
where capital indices are compound indices for the \textit{spin}-orbitals $p\sigma$ and $q\tau$; $\sigma$ and $\tau$ indicate the spin z-axis projection; $\theta_{pq}$ and $\theta'_{PQ}$ are rotation amplitudes; and the excitation operators are $\XX{P}{Q}=\CR{P}\AN{Q}$. Throughout this manuscript, lowercase letters indicate spatial orbitals and uppercase letters indicate spin-orbitals. Note that to have orbital, and not \textit{spin}-orbital, rotations, the $\theta'_{PQ}$ for $\alpha$ and $\beta$ spin orbitals must be equal and no spin flips are allowed; i.e.,  $\theta'_{p\alpha q\alpha}=\theta'_{p\beta q\beta}$ and $\theta'_{p\alpha q\beta}=\theta'_{p\beta q\alpha}=0$.

\textit{Construction:} Each of the planted solution classes have Hamiltonians $\hat H_\mathrm{sol}(\{\zeta_i\})$ defined with the parameters $\zeta_i$. To construct a planted solution instance, we choose values for both $\zeta_i$ and $\theta_{pq}$ or $\theta'_{PQ}$. There are at least three general ways of selecting these parameters: (i) choose the parameters arbitrarily, (ii) extract the parameters from an extant Hamiltonian, and (iii) optimize over the parameters to fit the planted solution to an extant Hamiltonian. For this latter case, we choose the cost function as $|| U(\{\theta_{pq}\})^\dagger \hat H_\mathrm{sol}(\{\zeta_i\}) U(\{\theta_{pq}\}) - \hat H_{\rm mol}||_2$, where $||\cdot||_2$ represents the 2-norm of the one- and two-body tensors and $\hat H_{\rm mol}$ is the extant molecular Hamiltonian. Additional details useful for practical construction of the planted solutions are in Appendix~\ref{app:construction}.

\textit{One-electron term:} For planted solution classes that focus solely on the two-electron term, we combine the one-electron term with the two-electron term. We discuss the details of how to perform these combinations for both spin-orbitals and spatial orbitals in Appendix~\ref{app:one_el}.

{\it Symmetries}: For the planted solutions to be more realistic, they should have the same symmetries as real electronic Hamiltonians. Here, we consider non-relativistic real-valued electronic Hamiltonians without external electro-magnetic fields. We also assume that $\alpha$ and $\beta$ electrons are described with the same real-valued orbitals. Thus, the planted solution classes generally observe the following symmetries:
\begin{itemize}
    \item number of electrons, $\HNe$
    \item electron spin 
projection, $\HSz$
    \item total spin, $\HSS$
    \item permutation symmetries \cite{Helgaker}
    \begin{itemize}
        \item two-electron integral eightfold symmetry, e.g., $g_{pqrs}=g_{rspq}$, etc.
        \item one-electron integral twofold symmetry, \newline i.e.  $h_{pq}=h_{qp}$
    \end{itemize}
    \item spin exchange symmetry \cite{Helgaker},  i.e., for the two-electron tensor in spin-orbital notation $g_{p\sigma,q\sigma,r\tau,s\tau}$
    \begin{align}
        g_{p\alpha,q\alpha,r\alpha,s\alpha} &= g_{p\alpha,q\alpha,r\beta,s\beta} \nonumber\\
        &= g_{p\beta,q\beta,r\alpha,s\alpha} \nonumber\\
        &= g_{p\beta,q\beta,r\beta,s\beta}. \nonumber
    \end{align}
\end{itemize}

{\it Killer operators}: The above symmetries are not only useful for improving the physical realism of the planted solutions, but can be used to add obfuscation and to provide additional degrees of freedom.
Consider a planted solution Hamiltonian that has a specific set of 
symmetries $\{\hat S_i\}$, where $\hat S_i$ are operators that commute with the Hamiltonian. Each eigenstate of $\hat H_\mathrm{sol}$  resides within one or more of the symmetry subspaces defined by the symmetry operator eigenvalues $\{s_i\}$. Labelling an eigenstate $\ket{\{s_i\}}$ by its symmetries, we have:
\bea
\hat H_\mathrm{sol}\ket{\{s_i\}} &=& E_{\{s_i\}} \ket{\{s_i\}}\\
\hat S_i \ket{\{s_i\}} &=& s_i \ket{\{s_i\}}.
\eea

To exploit these symmetries, we use the killer operators \cite{loaiza_block_2023}
\bea\label{eq:Km}
\hat K &=&  \sum_i \hat O_i (\hat S_i-s_i\hat 1),
\eea
where the form of the $\hat O_i$ operators is selected to ensure that $\hat K$ is hermitian and acts on at most two-electrons.

By construction, each of the terms in the killer operator acts as zero on a state within the specified symmetry subspace \cite{loaiza_block_2023}: $\hat K \ket{\{s_i\}} = 0$. Given that the ground state is in a specific symmetry subspace, the killer operators can be constructed so they leave the ground-state block of the Hamiltonian unchanged. This allows us to modify much of the Hamiltonian without impacting the planted ground state or its energy.

Depending on the exact symmetry $\hat S_i$ used, the form of the killer operator shown in Eq.~\ref{eq:Km} may not respect the two-body tensor permutation symmetries. In this case, we define the killer operator as
\bea\label{eq:Km_symmetrized}
\hat K 
&=&  \frac{1}{2}\sum_i \left[\hat O_i (\hat S_i-s_i\hat 1) + (\hat S_i-s_i\hat 1)\hat O_i\right].
\eea
To ensure $\hat K \ket{\{s_i\}} = 0$, we impose the additional restriction that $\left[\hat O_i,\hat S_i\right]=0$. 

\textit{Balance operators}: Quantum chemistry methods typically only target a specified \textit{global} symmetry subspace (e.g., total number of electrons). However, it is important for some planted solution classes that the ground state observes certain \textit{local} symmetries (e.g., the number of electrons within a given set of orbitals). However, these local symmetries may not commute with orbital rotations and can be lost during obfuscation of the Hamiltonian structure. In addition, any killer operators based on \textit{local} symmetries can modify the spectrum within a specified \textit{global} symmetry subspace, such that an excited state may drop in energy below the original ground state.

To address these two concerns, we use a restricted form of the killer operators to define the balance operators:
\bea
\hat B = \sum_j b_j (\hat S_j - s_j \hat 1)^2, \label{eqn_balance_op}
\eea
where $b_j>0$ and $j$ indexes the targeted symmetries. As this operator is positive semidefinite, it raises the energy of all of the states not within the targeted symmetry subspace, whether this symmetry is local or global. 
This action allows the  correct ground state to be retained if the $b_j$ are appropriately adjusted.
These balance operators improve obfuscation and flexibility by adding degrees of freedom to the Hamiltonian. Similarly to the killer operators, these balance operators may need adjustment to respect permutation symmetries.

\textit{Final form}:
Given properly symmetrized killer operators $\hat K^\mathrm{(S)}$, balance operators  $\hat B^\mathrm{(S)}$, and orbital rotations $U$, the final form of a planted solution Hamiltonian is
\begin{align} 
    \hat H_\mathrm{PS} &= U^\dagger\left(\hat H_\mathrm{sol} + \hat K^\mathrm{(S)} + \hat B^\mathrm{(S)}\right)U \label{eq:final_ps} \\
    &= \sum_{pq}\tilde h_{pq}\EE{p}{q} + \sum_{pqrs}\tilde g_{pqrs}\EE{p}{q}\EE{r}{s},
\end{align}
where $\tilde h_{pq}$ and $\tilde g_{pqrs}$ are the final one- and two-electron integrals, respectively.
All planted solution instances discussed in this work take the form (\ref{eq:final_ps}), with each planted solution class defining a different form for $\hat H_\mathrm{sol}$.
 
{\it Adversarial attacks}: For each planted solution class, we  discuss possible ways of exploiting the planted solution structure to solve it more easily than intended; we refer to such exploitations as adversarial attacks (AAs). These AAs can come from two approaches: (i) finding orbital rotations $U$ that reveal the underlying structure of the planted solution Hamiltonian and its matrix elements; and, (ii) taking a variational approach with  wavefunction ansatze that well match the structure of the ground state. We term these as structure-revealing AAs and variational-ansatz AAs, respectively.

However, the killer and balance operators  hinder the structure-revealing AAs. These operators increase the number of parameters that need to be optimized over and complicate the operator structure to be revealed. While the number of parameters is polynomial 
in $N$ (${\sim}N^2$ amplitudes from the orbital rotations and ${\sim}N^2$ parameters from the killer and balance operators), the problem is in general neither convex nor linear. Given that quadratic optimization is already NP-hard when the problem-defining matrix has a single negative eigenvalue \cite{pardalos_quadratic_1991}, optimization-based structure-revealing AAs are also in general NP-hard. Lack of knowledge of which symmetry generators $\hat{S}_i$ were used in the killer and balance operators further hinders these AAs.

\subsection{Mean-field solvable Hamiltonians}
\label{sec:HFsol}

One of the simplest exactly solvable Hamiltonians $\hat H_\mathrm{sol}$ are the mean-field Hamiltonians $\hat H_{\rm MF}$ that are diagonal in the 
Slater-determinants basis:  
\bea\label{eq:HFs}
\hat H_{\rm MF} =  \sum_{PQ} \lambda_{PQ} \hn_P \hn_Q, 
\eea
where $\hn_P = \XX{P}{P}$ and $\lambda_{PQ}\in \mathbb{R}$ for spin-orbitals $P$ and $Q$. 

Calculation of the ground-state energy of such a Hamiltonian is a quadratic unconstrained optimization problem (QUBO), which is known to be NP-hard in general \cite{barahona_computational_1982}. However, the problem sizes we could reasonably expect to encounter are up to about 300 binary variables. Problems of this size can be expected to be solvable within about an hour \cite{mittelmann_benchmark_2025,mittelmann_qubo_qplib_2025}. Thus, standard solvers such as IBM CPLEX \cite{cplex}, the Gurobi Optimizer \cite{gurobi}, or the open-source Solving Constraint Integer Programs (SCIP) \cite{BolusaniEtal2024OO} can be expected to obtain the solutions easily.

{\it Symmetries:} 
The number of electrons $\hat N_e$ is conserved 
by construction. The spin symmetries $\hat S_z$ and $\hat S^2$ are also conserved if 
the $\alpha$ and $\beta$ spin-orbital parameters 
in $\hat U$ and $\lambda_{PQ}$ are set to be equal and spin flips are forbidden.

{\it Adversarial attacks:} 
The ability to solve the bare $\hat H_{\rm MF}$ with standard solvers in small amounts of time also makes the obfuscated instance vulnerable to a variational-ansatz AA based on a single Slater determinant. Furthermore, if $\lambda_{PQ}$ is diagonal, the obfuscated Hamiltonian is solvable by Hartree-Fock \cite{Helgaker} in $O(N^4)$ time per iteration \cite{sousa_general_2007}. 

\subsection{Linear combination of commuting fermionic operators}
\label{sec:licop_hamiltonians}

The mean-field solvable Hamiltonians of Sec.~\ref{sec:HFsol} are built from sets of number operators that mutually commute. This mutual commutativity makes the number operators simultaneously diagonalizable by a single unitary operator.
Here, we expand beyond the number operators to a more general set of mutually commuting one- and two-electron operators. This expansion leads to the LInear combination of Commuting fermionic  OPerators (LICOP) planted solution class. This class includes operators of the form $\hat{G}_{Q}^{P} = \frac{1}{2}\left(\XX{P}{Q} + \XX{Q}{P}\right)$ 
and $\hat{G}_{QS}^{PR} = \frac{1}{8}\mathrm{S_8}\!\left(\XX{P}{Q}\XX{R}{S}\right)$, where $\mathrm{S_8}\!\left(\XX{P}{Q}\XX{R}{S}\right)$ indicates the positive sum of all eight permutations of $PQRS$ that correspond to the eightfold symmetry of the two-body tensor \cite{Helgaker}. 

To build a given instance of this planted solution class, we choose a subset of operators $\{\hat{G}_i\}$ from $\{\hat{G}_{Q}^{P}, \hat{G}_{QS}^{PR}\}$, with $[\hat{G}_j, \hat{G}_k] = 0 , \forall \hat{G_j},\hat{G_k} \in \{\hat{G}_i\}$. However, simultaneous diagonalizability of the $\hat G_i$ operators is not sufficient for the planted solution to be viable as the ground-state energy must be obtainable at reasonable cost. 
We can obtain the ground-state energy using only Clifford transformations, and thus in polynomial time on a classical computer \cite{nielsen_quantum_2010}, if all of the Pauli terms generated by the Jordan-Wigner transformation of the set $\{\hat G_i\}$ mutually commute \cite{bettaque_structure_2024,Yen2019b}. 

Through inclusion of the twofold and eightfold symmetries in the definition of $\hat G_i$, the Pauli terms generated by a \textit{single} $\hat G_i$ operator mutually commute by construction. To ensure the mutual diagonalizability of Pauli terms originating from \textit{different} $\hat G_i$ operators, we can (i) choose operators that act on disjoint sets of orbitals or (ii) choose operators that share a common repeated index, such as $\hat{G}_{QR}^{PR}$ and $\hat{G}_{UR}^{TR}$. Appendix~\ref{App:Comm} contains the mathematical derivations for both the single-operator and multi-operator cases.
With such choices for the operators, the procedure given in Appendix~B of Ref.~\citenum{Yen2019b} finds the required Clifford transformations in $O(N^3)$ time.
These transformations diagonalize the Hamiltonian
\begin{align}
    \hat H_\mathrm{LICOP} = \sum_{i}\lambda_{i}\hat{G}_{i}, \label{eqn_licop_hami}
\end{align}
where $\lambda_i \in \mathbb{R}$. 

{\it Symmetries:} By construction, $\hat H_\mathrm{LICOP}$ commutes with the number operator $\hat N_e$. The spin symmetries $\hat S_z$ and $\HSS$ are preserved if spin flips are prohibited and $\alpha$ and $\beta$ electrons are treated identically. 

{\it Adversarial attacks: } By expanding the set of admissible operators, we overcome the susceptibility of the mean-field solvable Hamiltonians to being solved with a single Slater-determinant ansatz. Instead, a variational-ansatz AA against a LICOP planted solution requires optimization over the Clifford group. This is computationally prohibitive as the group is discrete and exponentially large in $N$. 

A structure-revealing AA could attempt to
find a unitary rotation that, after conjugation of $\hat H_\mathrm{LICOP}$,
produces a polynomial of mutually-commuting fermionic operators. However, finding such a 
$U$ would be difficult as there are $O(N^8)$ possible commutators between all one- and two-electron fermionic operators. This AA is further thwarted with killer and balance operators that do not commute with $\hat H_\mathrm{LICOP}$. 

\subsection{Complete-active-space solvable Hamiltonians}
\label{sec:CASsol}

The complete-active-space solvable (CASS) planted solution class is structured around separable Hamiltonians $\hat H_\mathrm{SEP}$ whose subsystems $\{I\}$ are sufficiently small that they are exactly solvable by full configuration interaction in a reasonable time. The Hamiltonian structure is
\bea
\label{eq:CAS_new}
\hat H_\mathrm{SEP} &=& \sum_I \hat H_{\rm CAS}^{(I)} \label{eqn_h_cas_total} \\
\hat H_{\rm CAS}^{(I)} &=& \sum_{pq \in S_I} \lambda_{pq} \EE{p}{q} + \sum_{pqrs \in S_I} \lambda_{pqrs} \EE{p}{q} \EE{r}{s}, \label{eqn_h_cas_part}
\eea
where $S_I$ is the set of spatial orbitals $pqrs$ contained within subsystem $I$. All of the $S_I$ are disjoint and their union is the set of all spatial orbitals for the system.

We can obtain the exact ground-state energy $E_I$ and wavevector $\ket{\Psi_I}$ of each subsystem in $O(n_R^3)$ time, where $n_R = \max_{I} \binom{2N_{I}}{N_{e}^{(I)}}
         $ is the largest basis size of all subsystems $I$, ${N_I}$ is the number of orbitals in the subsystem, and $N_{e}^{(I)}$ is the number of electrons. The total ground-state energy and wavevector are then
\bea
E_\mathrm{GS} &=& \sum_I E_I, \\
\label{eq:CAS_gs}
\ket{\Psi_\mathrm{GS}} &=& \otimes_I \ket{\Psi_I}.
\eea

However, for the Hamiltonian as defined in Eq.~\ref{eq:CAS_new}, obtaining $E_\mathrm{GS}$ is complicated by the large number of possible partitions of the electrons between the subsystems and the possible spin states of each subsystem. To address this, we add balance operators that target the number of electrons within each subsystem $N_{e}^{(I)}$. These balance operators increase the energy of all states without the correct number of electrons in a given subsystem.
We then only allow an even number of electrons in each subsystem and specify that the desired ground state of each subsystem, and thus the whole system, is the singlet state. 
If, for a given planted solution, the above is not sufficient to maintain the singlet nature of the ground state after orbital rotation, balance operators based on the total spin and/or spin projection of each subsystem can be added to enforce the singlet nature. 

With the above restrictions on the electron partitioning and the spin states, the total ground-state energy and the total eigenstates require  $O(n_In_R^3)$, where $n_I\sim N/N_{I}$ is the number of subsystems. 

{\it Symmetries:}
By construction, these CASS planted solutions respect the symmetries $\hat N_e$, $\hat S_z$ and $\hat S^2$. Each of these symmetries have local counterparts that are present \textit{within} each subsystem: $\hat N_{e}^{(I)}$, $\hat S_z^{(I)}$ and $\hat S^{2(I)}$.

{\it Adversarial attacks:} 
Without knowledge of the orbital partition and the distribution of the electrons, the optimization space for either a structure-revealing or a variational-ansatz AA will include discrete variables and be combinatorially large. This is an NP-hard problem \cite{belotti_mixedinteger_2013}. Even with knowledge of the orbital partition and the distribution of the electrons, orbital rotations alone cannot be used to reveal the matrix elements of each subsystem as a Hamiltonian with killer and balance operators can easily be made non-separable, without changing the ground state or its energy. Furthermore, a product state variational ansatz based on the orbital partition and electron distribution is not useful; the ground state of the Hamiltonian in the obfuscated basis is not a product of states defined over \textit{disjoint} sets of orbitals.

\subsection{Seniority-conserving Hamiltonians}
\label{sec:seniority_conserving}

In this section, we consider exactly solvable Hamiltonians constructed from fermionic pair operators that form an $su(2)$ Lie algebra within a given spatial orbital $p$:
\begin{align}
    \hat{S}_p^+ &= \hat{a}_{p\uparrow}^\dagger \hat{a}_{p\downarrow}^\dagger\\
    \hat{S}_p^{-} &= \hat{a}_{p\downarrow} \hat{a}_{p\uparrow}\\
    \hat{S}_{p}^{z} &= (\hat{n}_p - 1)/2
\end{align}
where $\hat{n}_p$ counts electrons in spatial orbital $p$:
\begin{equation}
    \hat{n}_p = \hat{n}_{p\uparrow} + \hat{n}_{p\downarrow}.
\end{equation}
The seniority operator $\hat{\Omega}_p$ counts the number of unpaired electrons in spatial orbital $p$, and the total seniority operator $\hat{\Omega}$ counts the total number of unpaired electrons
\begin{align}
    \hat{\Omega}_p &= \hat{n}_{p\uparrow} + \hat{n}_{p\downarrow} - 2 \hat{n}_{p\uparrow}\hat{n}_{p\downarrow}\\
    \hat{\Omega} &= \sum_p \hat{\Omega}_p.
\end{align}
Note that operators $\hat{S}_p^\pm, \hat{S}_p^z$ 
act as zero on any state $\ket{\psi}$ with an unpaired electron in spatial orbital $p$. Conversely, states with an unpaired electron on the $p^{\rm th}$ orbital satisfy $\hat{\Omega}_p \ket{\psi} = \ket{\psi}$. 

\subsubsection{Antisymmetrized Geminal Power}
\label{sec:agp_hamiltonian}

These Hamiltonians are constructed to be eigen-Hamiltonians of antisymmetrized-geminal-power (AGP) states \cite{colemanStructureFermionDensity1965}. AGP states are alternative mean-field reference wavefunctions that are more accurate at modeling strong correlation than Hartree-Fock, while conserving the same particle number and spin symmetries \cite{hendersonGeminalbasedConfigurationInteraction2019,hendersonCorrelatingAntisymmetrizedGeminal2020}. Defined up to normalization, the AGP state is:
\begin{equation}
    \ket{\text{AGP}} = \left(\hat{\Gamma}^\dagger\right)^{N_\text{pair}}\ket{0},
\end{equation}
where $\ket{0}$ is the vacuum state with no electrons, $N_\text{pair}$ is the number of electron pairs ($N_\text{pair} = N_e/2$ for even number of $N_e$), and $\hat{\Gamma}^\dagger$ is a geminal creation operator:
\begin{equation}
    \hat{\Gamma}^\dagger = \sum_{p=1}^{N} c_p \hat{S}_p^+. \label{eqn:geminal_creation_op}
\end{equation}
The state $\ket{\rm AGP}$ expands into a combinatorial number of Slater determinants with coefficients parametrized by only $N$ constants $c_p\in \mathbb{R}$, where $N$ is the number of orbitals.

To construct a Hamiltonian which $\ket{\text{AGP}}$ acts as a ground state, we rely on the existence of one-electron killer operators $\hat{K}_{pq}$  for $\ket{\text{AGP}}$, defined as \cite{weinerExcitationOperatorsAssociated1983,hendersonCorrelatingAntisymmetrizedGeminal2020}:
\begin{equation}
    \hat{K}_{pq} = c_p\left(\hat{a}_{p\uparrow}^\dagger \hat{a}_{q\uparrow} + \hat{a}_{p\downarrow}^\dagger \hat{a}_{q\downarrow}\right) - c_q\left(\hat{a}_{q\uparrow}^\dagger \hat{a}_{p\uparrow} + \hat{a}_{q\downarrow}^\dagger \hat{a}_{p\downarrow}\right),
\end{equation}
where each of the $c_p$ and $c_q$ are the same as those in Eq.~\ref{eqn:geminal_creation_op}. The $\hat{K}_{pq}$  satisfy $[\hat{K}_{pq}, \hat{\Gamma}^\dagger] = 0$, such that $\hat{K}_{pq} \ket{\text{AGP}} = 0$. 
We form Hamiltonians using a linear combination of these killer operators $\tilde{K}_u = \sum_{p>q} d_{pq}^{(u)} \hat{K}_{pq}$ as follows:
\begin{align}
    \hat{H}_\text{AGP} &= \sum_u \omega_u \tilde{K}_u^\dagger \tilde{K}_u \label{eqn_agp_hami} \\
    &= \sum_{p > q} \sum_{r > s} \left[\sum_u \omega_u d_{pq}^{(u)} d_{rs}^{(u)}\right] \hat{K}_{pq}^\dagger \hat{K}_{rs}, \label{agp_expanded}
\end{align}
 where $\omega_u \in \mathbb{R}$, $\omega_u \geq 0$, and $d_{pq}^{(u)}\in\mathbb{R}$. The $\ket{\text{AGP}}$ are ground states of $\hat{H}_\text{AGP}$ as $\hat{H}_\text{AGP}\ket{\text{AGP}} = 0$ and $\hat{H}_\text{AGP}$ is a linear combination of positive operators $\tilde{K}_u^\dagger \tilde{K}_u$ with positive coefficients $\omega_u$. This ensures that all eigenvalues of $\hat{H}_\text{AGP}$ are non-negative. 

{\it Symmetries:} $\hat{H}_\text{AGP}$ conserves the total number of electrons $\hat N_e$, the spin symmetries $\hat S_z$ and $\HSS$, the total seniority operator $\hat{\Omega}$, and the local seniority operators $\hat{\Omega}_p$.

{\it Adversarial attacks:}
Based on the construction of the Hamiltonian, the ground state has an energy of zero by definition. Thus, merely knowing that the Hamiltonian is an AGP planted solution would reveal the ground-state energy. To mitigate this problem, we can  (1) add the terms $a \hat N_e$ and $b\hat N_e^2$ to shift the Hamiltonian spectrum of the electron number subspace uniformly and (2) add killer operators $\hat K =  \sum_{pq} c_{pq}\EE{p}{q} (\hat N_e-n_e\hat 1)$ where $n_e$ is the number of electrons of the desired ground state and the coefficients $c_{pq}\in\mathbb{R}$ are chosen randomly.  The randomized killer operators impede attempts to obtain $a$ and $b$ through linear fits of the operators $a \hat N_e$ and $b\hat N_e^2$ to collections of energy expectation values.

A variational AA over the space of AGP states $U(\vec{\theta})\ket{\text{AGP}(\vec{c})}$, where $\vec\theta$ are orbital rotation angles and $\vec{c}$ are the parameters defining the AGP coefficients, could still be used to obtain the ground-state energy. However, this requires a nonlinear, nonconvex optimization.

\subsubsection{Pairing Hamiltonians}
\label{sec:Psol}

The Pairing Hamiltonian $\hat{H}_{\text{Pair}}$, also called the reduced Bardeen-Cooper-Schrieffer (BCS) Hamiltonian, takes the following form \cite{richardsonRestrictedClassExact1963, Dukelsky_RG_Review_2004, JohnsonRG2020}:
\begin{equation}
    \hat{H}_{\text{Pair}} = \sum_{p=1}^{N} \epsilon_p \hat{n}_p - g\sum_{p,q=1}^{N} \hat{S}_p^{+} \hat{S}_q^{-}, \label{eqn_pairing_hami}
\end{equation}
where $\epsilon_p$ are single-particle energies and $g \in \mathbb{R}$ is the constant pairing length. The eigenstates of $\hat{H}_{\text{Pair}}$, called Richardson-Gaudin (RG) states, can be obtained in polynomial time and have been used as an approximate solution to an \textit{ab initio} electronic Hamiltonian \cite{JohnsonRG2020}. Using geminals of the following form
\begin{equation}
    \hat{\Gamma}_{\text{RG}}^\dagger (u) = \sum_{p=1}^{N} \frac{1}{u - \epsilon_p}\hat{S}_p^+,
\end{equation}
defined in terms of a complex parameter $u$ (a rapidity), the RG states can be constructed as follows:
\begin{equation}
    \ket{\text{RG}(\vec{u}, \xi)} = \prod_{\mu=1}^{N_\text{pair}} \hat{\Gamma}_{\text{RG}}^\dagger (u_\mu) \ket{\xi}.\label{eqn:rg_state}
\end{equation}
The initial state $\ket{\xi}$ is an eigenstate of all orbital seniority operators $\hat{\Omega}_p$, and contains no paired electrons. The orbitals $G$ that contain unpaired electrons are ignored by the geminals. Such a state is an eigenstate of $\hat{H}_{\text{Pair}}$ provided that the rapidities satisfy a set of nonlinear equations called Richardson's equations \cite{JohnsonRG2020}
\begin{equation}
    \frac{2}{g} + \sum_{p\not \in G} \frac{1}{u_\mu - \epsilon_p} + \sum_{\nu\not=\mu}^{N_{\text{pair}}} \frac{2}{u_\nu - u_\mu} = 0, \mu = 1,\ldots,N_\text{pair}. \label{eqn_rg_constraints}
\end{equation}
These equations can be solved in polynomial time \cite{faribaultGaudinModelsSolver2011,claeysEigenvaluebasedMethodFormfactor2015,JohnsonRG2020}. The corresponding eigenvalue of the RG state is $E_\mathrm{RG}(\vec{u}, \vec{\epsilon})=2\sum_{\mu=1}^{N_{\text{pair}}} u_\mu + \sum_{p\in G} \epsilon_p$, where $\vec{\epsilon}=\{\epsilon_p\}$.

{\it Symmetries:} $\hat{H}_\text{Pair}$ conserves the total number of electrons $\hat N_e$, the spin symmetries $\hat S_z$ and $\HSS$, and the total seniority operator $\hat{\Omega}$.

\textit{Adversarial attacks:}
A variational AA could form the ansatz from Eq.~\ref{eqn:rg_state}, subject to constraints of Eq.~\ref{eqn_rg_constraints}, and orbital rotations. However, this would require a nonconvex, nonlinear optimization.

\renewcommand{\arraystretch}{1.5} 
\begin{table*}
    \centering
    \caption{Summary of planted solution classes.  Each of the classes 1--5 define a different form for the exactly solvable Hamiltonian $\hat H_\mathrm{sol}$ of the general planted solution $\hat H_\mathrm{PS}$. $U$ are orbital rotations. $\hat K^\mathrm{(S)}$ and $\hat B^\mathrm{(S)}$ are symmetrized killer and balance operators, respectively. The remaining symbol definitions are in the text. Numbers in brackets refer to corresponding equations.}\label{tab_planted_solution_summary}
    \begin{tabular}{lll}
       
       {General planted solution (\ref{eq:final_ps})}:  & {\small $\hat H_\mathrm{PS} = U^\dagger\left(\hat H_\mathrm{sol} + \hat K^\mathrm{(S)} + \hat B^\mathrm{(S)}\right)U$~~~} & {Sec.~\ref{sec:general_features}}\\
       &&\\
       Planted Solution Class  & Form of $\hat H_\mathrm{sol}$ & Location\\
       \hline
       1. Mean-field solvable (\ref{eq:HFs})  & $\hat H_{\rm MF} =  \sum_{PQ} \lambda_{PQ} \hn_P \hn_Q$ & Sec.~\ref{sec:HFsol} \\
       2. LICOP (\ref{eqn_licop_hami}) & $\hat H_\mathrm{LICOP} = \sum_{i}\lambda_{i}\hat{G}_{i}$ & Sec.~\ref{sec:licop_hamiltonians}\\
       
       3. CASS (\ref{eqn_h_cas_total})  &    $\hat H_\mathrm{SEP} = \sum_I \hat H_{\rm CAS}^{(I)}$
        & Sec.~\ref{sec:CASsol}\\
       4. Antisymmetrized geminal power (\ref{eqn_agp_hami})  & $\hat{H}_\text{AGP} = \sum_u \omega_u \tilde{K}_u^\dagger \tilde{K}_u$ & Sec.~\ref{sec:agp_hamiltonian}\\
       5. Pairing (\ref{eqn_pairing_hami}) & 
     $\hat{H}_{\text{Pair}} = \sum_{p=1}^{N} \epsilon_p \hat{n}_p - g\sum_{pq=1}^{N} \hat{S}_p^{+} \hat{S}_q^{-}~~~$
     & Sec.~\ref{sec:Psol}\\
    \end{tabular}
\end{table*}

\section{Demonstration for Cropped CASS Planted Solutions}
\label{sec:results}
Here, we construct and demonstrate several planted solution instances based on the CASS class discussed in Sec.~\ref{sec:CASsol}.
The one- and two-electron tensors of these planted solutions are based on two sets of industrially relevant systems: Mn mono and bridged dimolybdenum homogeneous catalysts \cite{qb_benchmark_repo,gsee_benchmark_paper,bellonziFeasibilityAcceleratingHomogeneous2024}. Mn mono is a water-oxidation catalyst studied in the context of CO${}_2$ reduction \cite{Crandell2017}. We examine three Hamiltonians from its high spin S=5/2, antiferromagnetically coupled, O-O bond formation pathway \cite{Crandell2017}, with 30-31 spatial orbitals in the active space \cite{qb_benchmark_repo,gsee_benchmark_paper}. The bridged dimolybdenum complex is a nitrogen fixation catalyst \cite{bellonziFeasibilityAcceleratingHomogeneous2024, Tanaka2014}. We examine three Hamiltonians that represent the reactant, transition, and product states of the reaction pathway. These Hamiltonians have 69-70 spatial orbitals in the active space. The Hamiltonians for these systems were previously defined within their active spaces using the atomic valence active space (AVAS) method \cite{bellonziFeasibilityAcceleratingHomogeneous2024, sayfutyarova_automated_2017} and can be found in the QB-GSEE Benchmark \cite{qb_benchmark_repo,gsee_benchmark_paper} GitHub repository. The corresponding unique universal identifiers (UUIDs) are listed in Appendix~\ref{app:UUIDs}.

To build the CASS planted solutions from these six Hamiltonians, we (i) choose an orbital partition $\{S_I\}$ to define the subsystems and (ii) compose the one- and two-electron tensors of each subsystem from cropped versions of the tensors of the source Hamiltonian. The cropped tensors are the original tensors with any terms that would otherwise couple the subsystems removed.
The final planted solutions include symmetrized killer $\hat K^\mathrm{(S)}$ and balance operators $\hat B^\mathrm{(S)}$ and follow the form of Eq.~\ref{eq:final_ps}:
\begin{align} 
    \hat H_\mathrm{PS} = U^\dagger(\vec\theta)\left(\sum_I \hat H_{\rm CAS}^{(I)} + \kappa\hat K^\mathrm{(S)} + \hat B^\mathrm{(S)}\right)U(\vec\theta), \label{eq:cropped_cass_hami}
\end{align}
where we sum over the orbital subset index $I$; the orbital subset size $|S_I|=N_\mathrm{s}$, except for the final partition where $\left|S_I\right|\leq N_\mathrm{s}$; $N_\mathrm{s}$ is the upper bound on the subsystem size; $\kappa$ is the scaling value of the killer operators;  $\vec\theta=\{\theta_i\}=\mathrm{RNG}(\Theta,S_\mathrm{RNG})$, such that $\theta_i\in[0,\Theta]$; $\Theta$ defines the allowed range of the orbital rotation parameters; and $S_\mathrm{RNG}$ is the seed of the random number generator (RNG). The planted solution construction code is stored in a public GitHub repository, see Appendix~\ref{app:code}. For the initial set of planted solutions, $N_\mathrm{s}=4$,  $\kappa=0.01$, and $\Theta=0.1$.

The convergence behaviour of DMRG calculations as a function of the bond dimension provides information on the difficulty of a planted solution or other Hamiltonian.
We use DMRG \cite{whiteDmrg1992,whiteDmrg1993} as implemented in Block2 \cite{zhai2023block2} to estimate the ground-state energies of the planted solutions and their source Hamiltonians.  The bond dimension needed to achieve 1 mHa accuracy provides a quantitative measure of the solution difficulty, while the shape of the convergence curve provides qualitative insight.

Each series of DMRG calculations applied to the catalyst and planted solution Hamiltonians starts at a bond dimension of 4. The bond dimension is then increased by 10\% per step. This continues until the energy changes by $<0.05$ mHa or an overall 24h time limit is reached. These calculations are run on the Niagara cluster hosted by SciNet \cite{ponce2019deploying,loken2010scinet}; more computational details are in Appendix~\ref{app:UUIDs}.  The reference energies of the source Hamiltonians are based on extrapolated DMRG calculations previously performed in Ref.~\cite{bellonziFeasibilityAcceleratingHomogeneous2024}. 

\begin{figure*}[tp]
    \centering
    \includegraphics[width=0.75\textwidth]{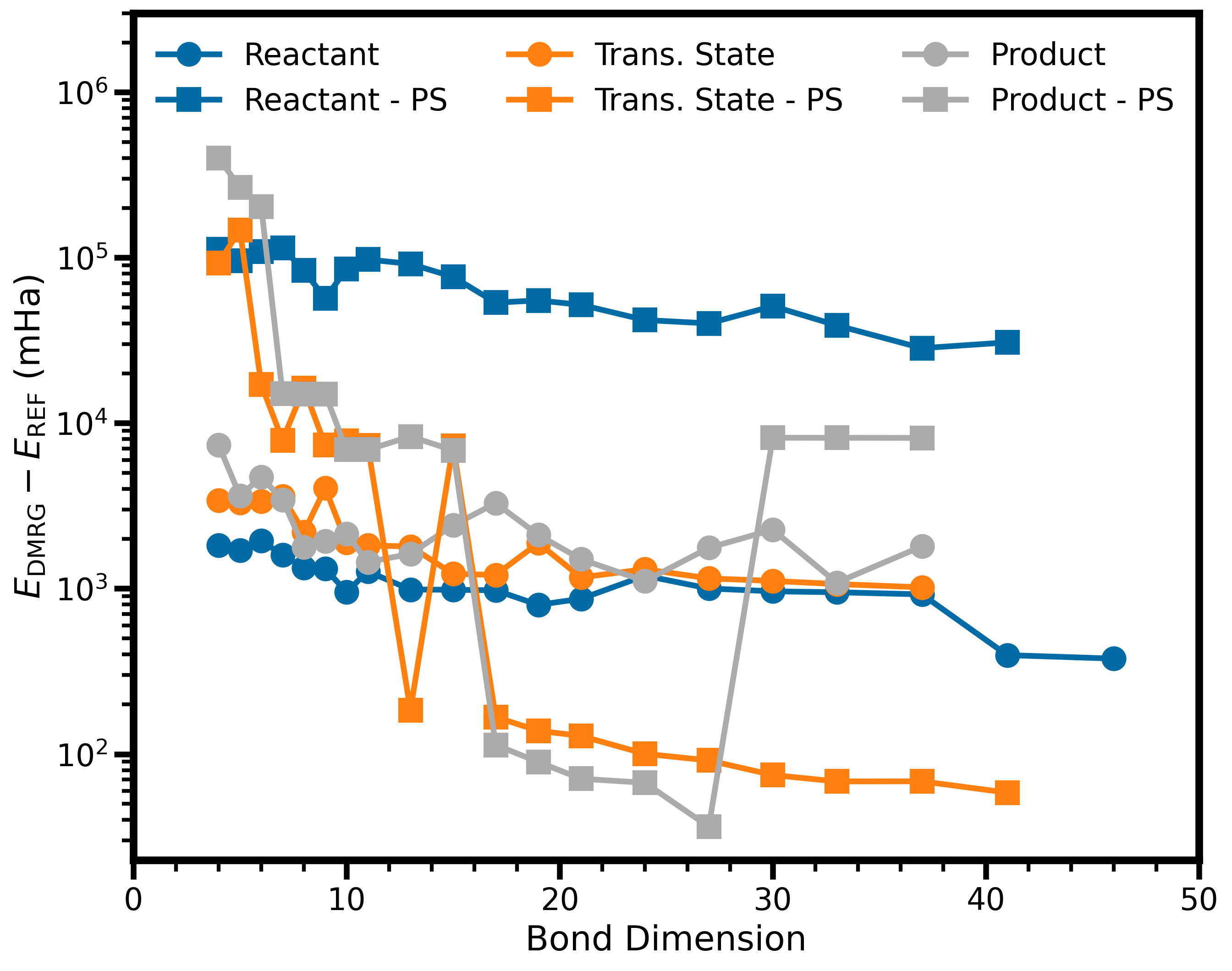}
    \caption{
    Convergence plot of DMRG calculations as a function of bond dimension, for the three bridged dimolybdenum catalyst reaction steps \cite{bellonziFeasibilityAcceleratingHomogeneous2024,Tanaka2014} (reactant, transition state and product; circles) and corresponding CASS planted solutions (PS; squares). The reactant and transition state have 69 spatial orbitals and the product has 70 spatial orbitals in the active space \cite{bellonziFeasibilityAcceleratingHomogeneous2024}. 
    }
    \label{fig:ccas_drmg_mo2n2_orig}
\end{figure*}

\begin{figure*}[tp]
    \centering
    \includegraphics[width=0.75\textwidth]{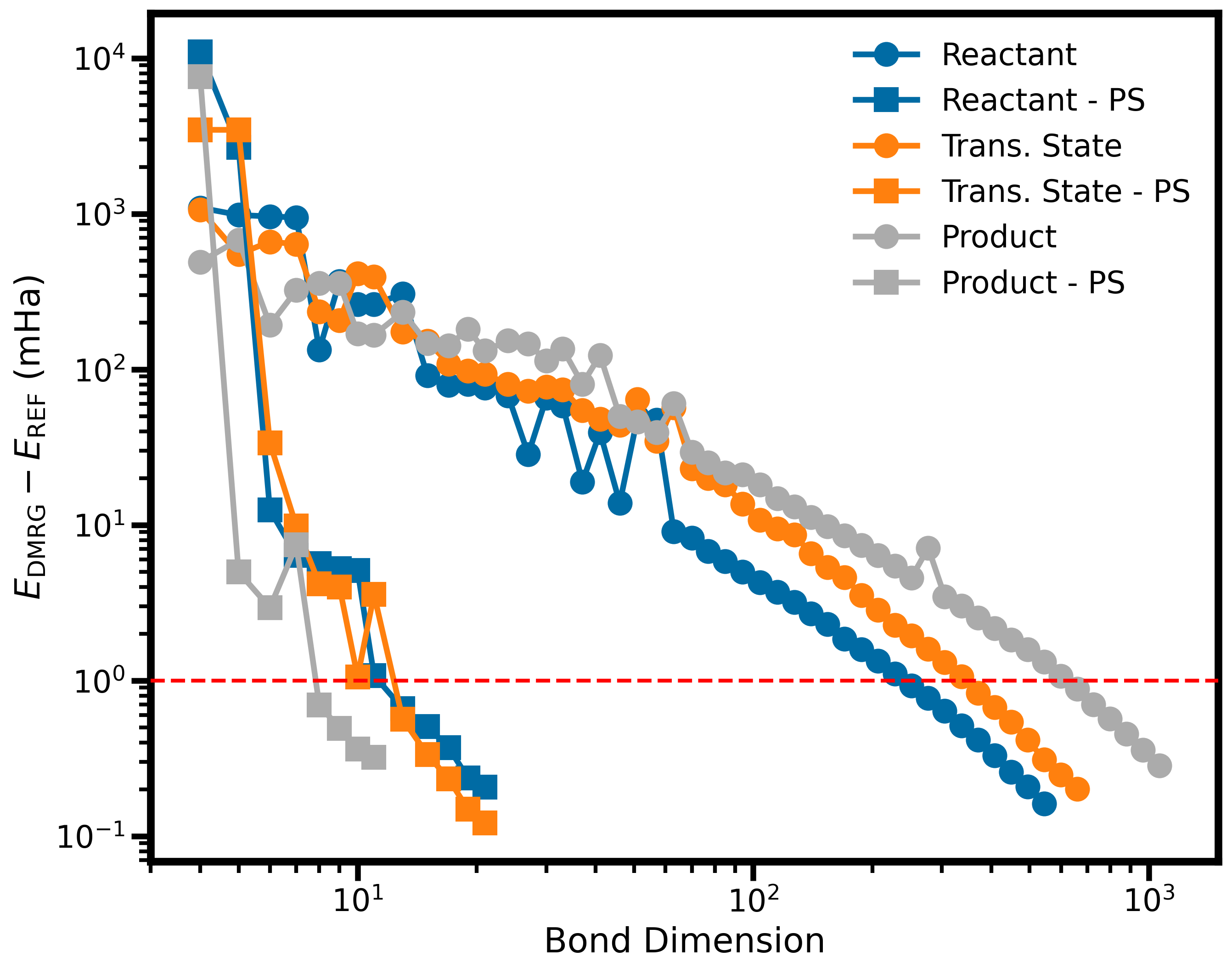}
    \caption{Convergence plot of DMRG calculations as a function of bond dimension, for the reactant (${}^5$15$_\mathrm{AF}$), transition state (${}^5$15$_\mathrm{AF}$-TS), and  product (${}^5$16) of the high spin S=5/2, antiferromagnetically-coupled,  O-O bond formation pathway of the Mn Mono catalyst (circles) \cite{Crandell2017}, with 31, 31, and 30 spatial orbitals in the active space \cite{bellonziFeasibilityAcceleratingHomogeneous2024}, respectively. Calculations for the corresponding CASS planted solutions (squares) are also shown.
    }
    \label{fig:ccas_drmg_mn_mono_orig}
\end{figure*}

Figure~\ref{fig:ccas_drmg_mo2n2_orig} shows that none of the DMRG calculation series for the bridged dimolybdenum catalysts or their planted solutions converge well to the reference energy; they end due to the 24 hr time limit. Based on the deviations from the reference energies, some of the planted solutions are harder for DMRG than their source Hamiltonians while others are easier, at least at the bond dimensions considered. The DMRG calculation for the product-based planted solution (grey squares) is also trapped in an excited state more often than the calculation for the product Hamiltonian itself (grey circles). 
In contrast to these results, Fig.~\ref{fig:ccas_drmg_mn_mono_orig} shows that the DMRG calculation series for the planted solutions based on the Mn mono catalyst converge significantly faster than the series for the original Hamiltonians. The original Hamiltonian calculations require bond dimensions of 200 or more to reach 1 mHa accuracy, while those for the planted solutions achieve this accuracy  with a bond dimension less than 20. 

From Figs.~\ref{fig:ccas_drmg_mo2n2_orig} and \ref{fig:ccas_drmg_mn_mono_orig}, it is clear that the difficulty to obtain the ground state of a planted solution can differ significantly from the  difficulty encountered with the corresponding source Hamiltonian. 
If the planted solutions are to have any practical use as part of a benchmark data set, however, it is important that their  difficulties be adjustable.

\begin{figure*}[tp]
    \centering
    \includegraphics[width=0.49\textwidth]{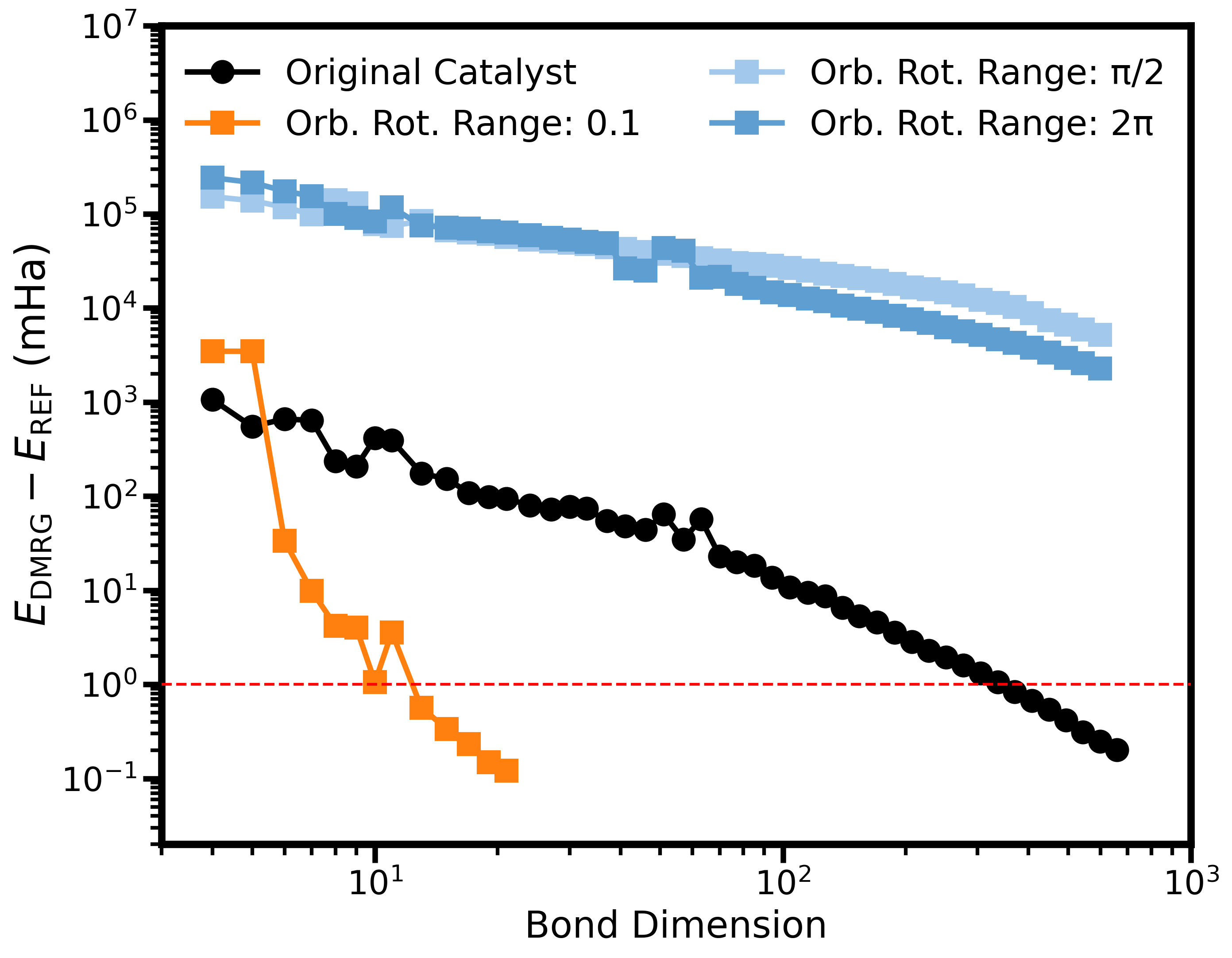}
    \includegraphics[width=0.49\textwidth]{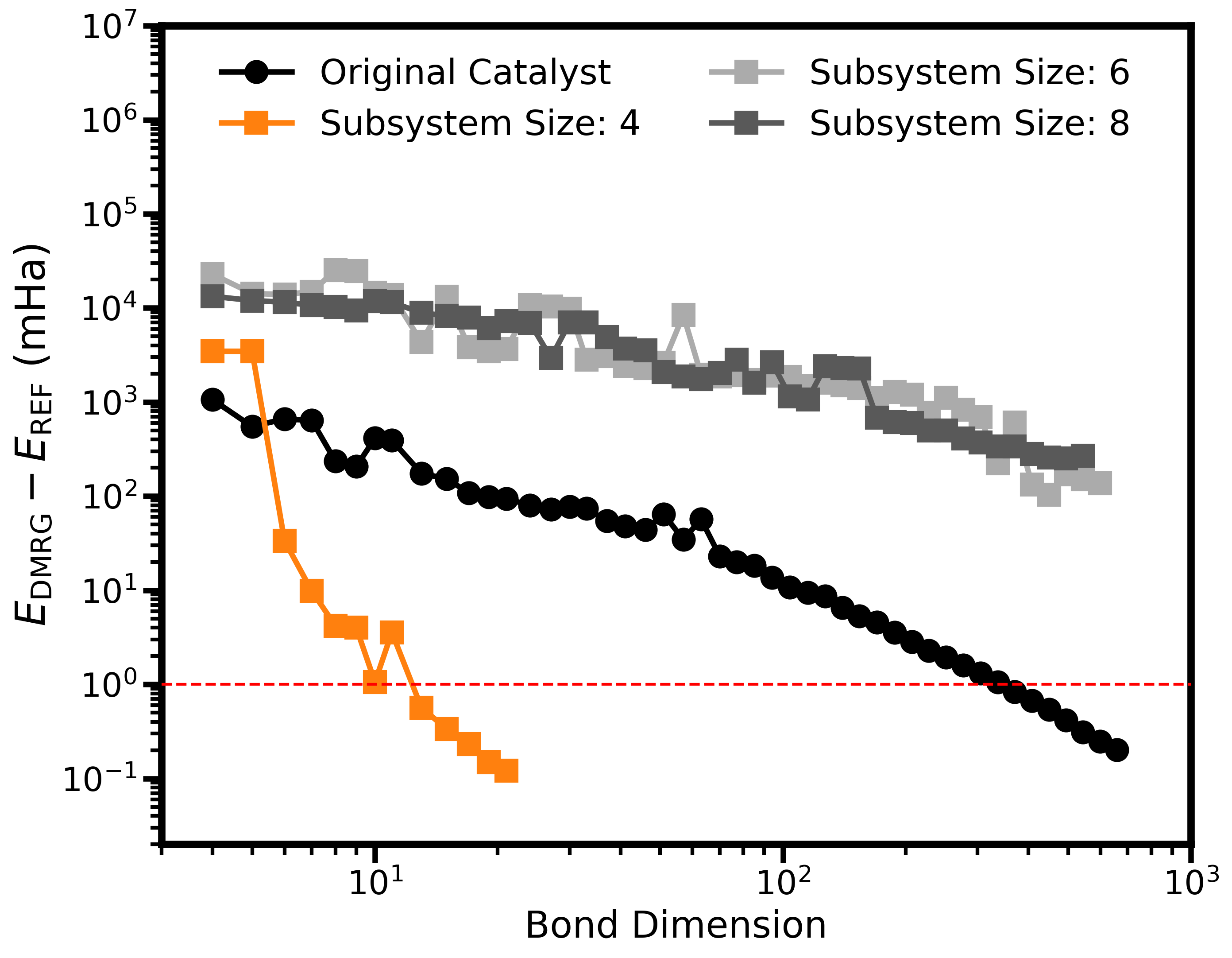}
    \includegraphics[width=0.49\textwidth]{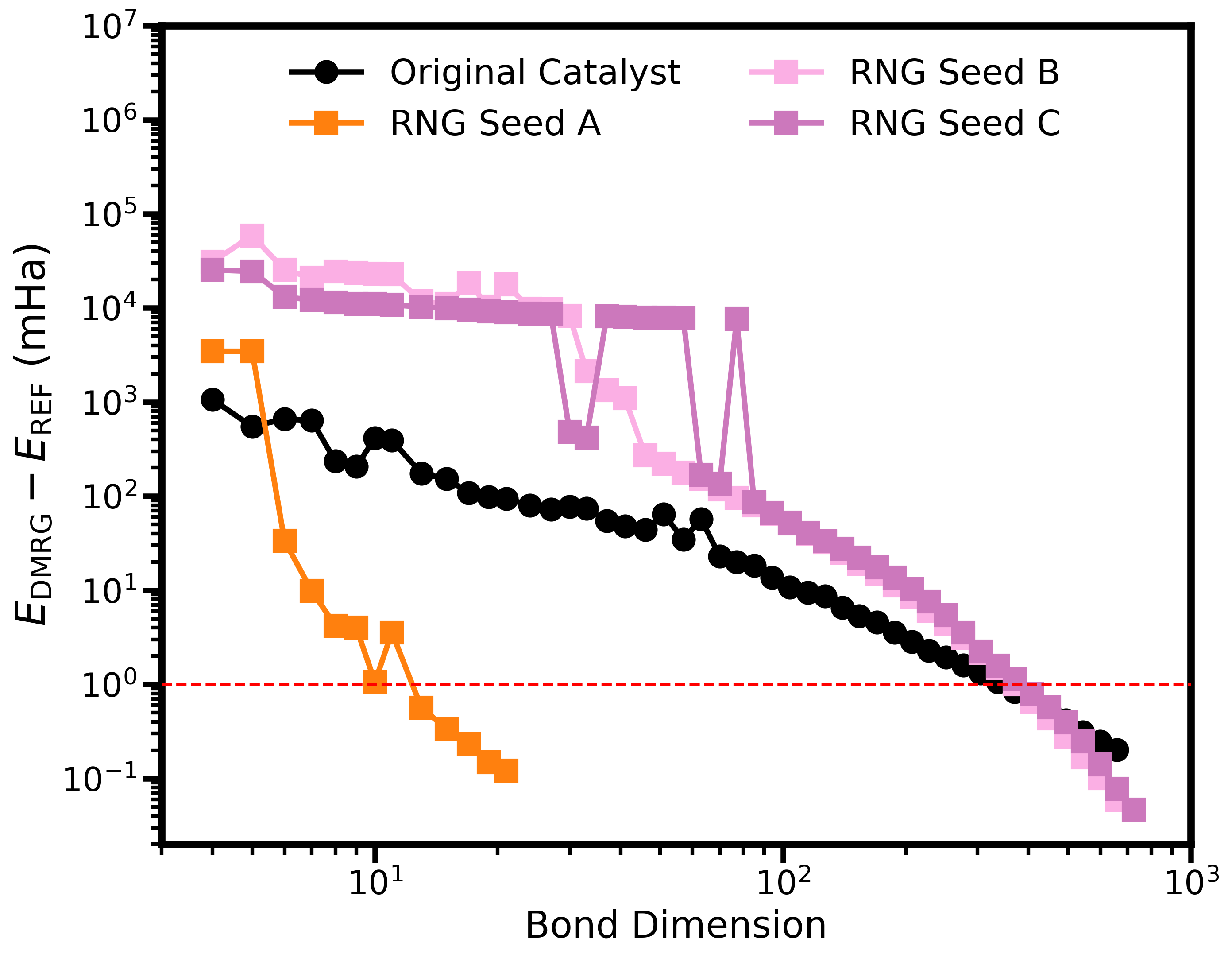}
    \includegraphics[width=0.49\textwidth]{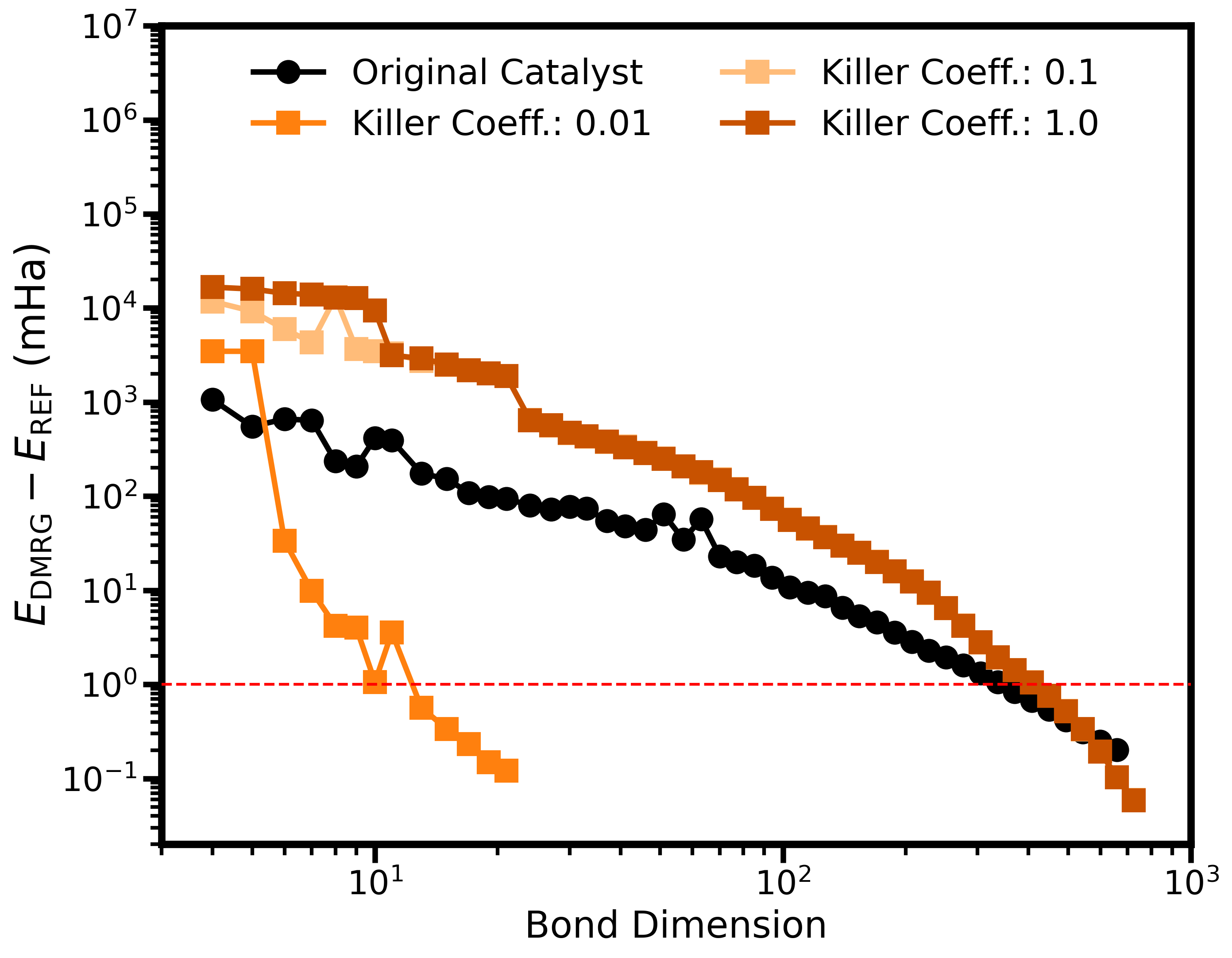}
    \caption{Convergence plots of DMRG calculations as a function of bond dimension for planted solutions with various construction parameters (squares), all derived from the transition state (${}^5$15$_\mathrm{AF}$-TS) Mn Mono catalyst Hamiltonian (black circles). The original planted solution, as previously shown in Fig.~\ref{fig:ccas_drmg_mn_mono_orig}, is in orange and has identical parameters for all subplots. Each subplot shows the variation of a single parameter. \textit{Upper left}: increase in orbital rotation range $\Theta$ (blues), \textit{upper right}: increase in maximum subsystem size $N_\mathrm{s}$ (greys), \textit{bottom left}: change in random number generator seed $S_\mathrm{RNG}$ (pinks), \textit{bottom right}: increase in killer coefficient $\kappa$ (reds). See Eq.~\ref{eq:cropped_cass_hami} for the definitions of these parameters. The dashed horizontal red line shows 1 mHa.}
    \label{fig:ccas_drmg_mn_mono_tuned}
\end{figure*}

To show this adjustment is possible, we build several planted solutions from the transition state of the Mn mono catalyst (${}^5$15$_\mathrm{AF}$-TS),  varying the construction parameters $N_\mathrm{s}$, $\kappa$, $\Theta$, and $S_\mathrm{RNG}$.
The DMRG calculations all use the same initial matrix product state. 
Figure~\ref{fig:ccas_drmg_mn_mono_tuned} shows the convergence curves for these different planted solutions. The planted solution Hamiltonians exhibit a wide range of difficulties, which vary from achieving an accuracy of 1 mHa with a bond dimension of less than 20 to not converging within the time limit.

Increasing the range  $\Theta$ of the orbital rotation parameters makes the planted solutions significantly harder than the original catalyst (\textit{upper left} subplot of Fig.~\ref{fig:ccas_drmg_mn_mono_tuned}). These orbital rotations change the basis of the Hamiltonian and have a chance of significantly changing the entanglement structure of the ground state, particularly for larger values of $\Theta$. The changing solution difficulty is thus in line with the known sensitivity of the efficiency of DMRG to the entanglement structure of the ground state \cite{wouters_density_2014}.
Increasing the size limit $N_\mathrm{s}$ of each subsystem from 4 to 6 or 8 (\textit{upper right} subplot of Fig.~\ref{fig:ccas_drmg_mn_mono_tuned}) also significantly raises the solution difficulty, though not to the same extent as that induced by the increase in $\Theta$. 
Increasing the killer coefficient $\kappa$ by a factor of 10 or 100 (\textit{lower right} subplot of Fig.~\ref{fig:ccas_drmg_mn_mono_tuned}) causes the planted solution difficulty to resemble the original catalyst. Furthermore, merely changing the seed $S_\mathrm{RNG}$ of the random number generator causes the calculations to converge at bond dimensions similar to those required for the original catalyst  (\textit{lower left} subplot of Fig.~\ref{fig:ccas_drmg_mn_mono_tuned}), while also increasing the likelihood the calculation is trapped within an excited state at various points along the convergence path. 

These results show that the solution difficulties of the planted solutions are quite sensitive to the construction parameters. Furthermore, three of the four parameters that we examined control the operators used for obfuscation. This implies that the obfuscation operators themselves contribute significantly to the solution difficulty.
Given these obfuscation operators are shared amongst all of the planted solution classes proposed in Sec.~\ref{sec:planted_sols_description}, it is expected that the solution difficulties of the other planted solution classes can be similarly adjusted. 

\section{Conclusion}
\label{sec:conclusion}

In this work, we developed several methods of generating planted solution Hamiltonians for the nonrelativistic electronic structure problem. These methods draw on sets of commuting fermionic terms, partition the system into a set of non-interacting subsystems, or employ pair operators. Symmetry-targeting killer operators, balance operators, and randomized orbital rotations are included to obscure the Hamiltonian structure without changing the ground state. We built several complete-active-space-solvable planted solutions based on industrially relevant catalyst Hamiltonians and examined the difficulty of calculating their ground-state energies (GSEs). This difficulty was assessed by considering the convergence behaviour of DMRG calculations as a function of bond dimension. We demonstrated that the difficulty of GSE calculations for planted solutions (1) varies relative to their source Hamiltonians and (2) can be significantly modified by changing the construction parameters of the planted solution.

These adjustable planted solutions provide a way of generating a significant number of large, chemically relevant, and difficult-to-solve Hamiltonians for which the ground-state energy is known. This is particularly useful for building benchmark data sets. These planted solutions also provide a set of tools useful for examining which Hamiltonian structures are difficult or easy for a particular GSE estimation method. Further, these planted solutions can be used to identify quantitative features of a Hamiltonian that correlate with its GSE calculation difficulty. Additional planted solution classes that correspond to spin-orbit-coupled systems, molecules in external magnetic fields, and relativistic systems can also be developed.

\section*{Acknowledgments}
A.F.I. gratefully acknowledges helpful discussions with Gustavo E. Scuseria, Norm M. Tubman, and Ilya G. Ryabinkin.
J.T.C. thanks A.M. Rey for her hospitality during his visit to JILA at CU Boulder.
Authors acknowledge financial support from the DARPA Quantum Benchmarking program and the Natural Sciences and Engineering Council of Canada (NSERC). This research was partly enabled by Compute Ontario (computeontario.ca) and the Digital Research Alliance of Canada (alliancecan.ca) support. Part of the computations were performed on the Niagara supercomputer at the SciNet HPC Consortium. SciNet is funded by Innovation, Science, and Economic Development Canada, the Digital Research Alliance of Canada, the Ontario Research Fund:~Research Excellence, and the University of Toronto. 
 
\appendix

\section{Code Availability}
\label{app:code}

The code for cropped CASS planted solutions can be found in \href{https://github.com/jtcantin/planted_solutions}{https://github.com/jtcantin/planted\_solutions}. \\

Code for generating planted solutions of various classes can be found in 
\href{https://github.com/iloaiza/planted_solutions}{https://github.com/iloaiza/planted\_solutions}. 

\section{One-electron components} \label{app:one_el}
Here, we discuss how the one-electron terms of the Hamiltonian can be combined with the two-electron terms. 
For spin-orbitals \cite{loaiza_reducing_2023},
\begin{align}
    \sum_{PQ} h_{PQ}\XX{P}{Q} &=  U \left[\sum_P \epsilon_P \hn_P \right]U^\dagger\nonumber \\
    &=U \left[\sum_P \epsilon_P \hn_P^2 \right]U^\dagger \nonumber \\ 
&=\left[\sum_P \epsilon_P (U \hn_P U^\dagger) 
(U \hn_P U^\dagger)\right] \nonumber \\
&= \sum_{PQRS} h_{PQRS} \XX{P}{Q}\XX{R}{S}, \label{eq:1to2}
\end{align}
where the last equality comes from $U \hn_P U^\dagger = \sum_{RS} b_{RS} \XX{R}{S}$. This approach relies on $\hn_P^2=\hn_P$.

For spatial orbitals, we can choose a definite number of electrons $N_e=\langle\sum_r \EE{r}{r} \rangle$ 
and then perform the following \cite{pyscf_absorb_h1e,Sun2020,Sun2018}: 
\begin{align}
H &= h_{pq}\EE{p}{q} + g_{pqrs}\EE{p}{q}\EE{r}{s} \nonumber \\
 &=\frac{1}{2}h_{pq}(\EE{p}{q}+\EE{p}{q}) + \frac{1}{2}g_{pqrs}\EE{p}{q}\EE{r}{s}\nonumber \\
 &=\frac{1}{2}h_{pq}(\EE{p}{q}\frac{\sum_r \EE{r}{r}}{N_e}+\frac{\sum_r \EE{r}{r}}{N_e}\EE{p}{q}) + \frac{1}{2}g_{pqrs}\EE{p}{q}\EE{r}{s}\nonumber \\
 &=\frac{1}{2N_e}(h_{pq}\delta_{rs}\EE{p}{q}\EE{r}{s}+h_{rs}\delta_{pq}\EE{p}{q}\EE{r}{s}) + \frac{1}{2}g_{pqrs}\EE{p}{q}\EE{r}{s}\nonumber \\
  &=\frac{1}{2N_e}(h_{pq}\delta_{rs}+h_{rs}\delta_{pq})\EE{p}{q}\EE{r}{s} + \frac{1}{2}g_{pqrs}\EE{p}{q}\EE{r}{s}\nonumber \\
  &=G_{pqrs}(N_e)\EE{p}{q}\EE{r}{s}
\end{align}
where $\sum_r \EE{r}{r}/N_e=1$ for all expectation values of states with $N_e$ electrons, $\sum_r$ is implicit after it is first introduced, $G_{pqrs}(N_e) = (h_{pq}\delta_{rs}+h_{rs}\delta_{pq})/(2N_e) + g_{pqrs}/2$, and repeated indices are summed.
As we only consider ground states with a specified number of electrons, we can always use this procedure for planted solution generation.

\section{Commutativity of Pauli products from commuting fermionic operators} \label{App:Comm}

To generate LICOP planted solutions as discussed in Sec.~\ref{sec:licop_hamiltonians}, all of the Pauli terms generated by the Jordan-Wigner transformation \cite{bettaque_structure_2024} of the $\hat{G}_{Q}^{P} = \frac{1}{2}\left(\XX{P}{Q} + \XX{Q}{P}\right)$ 
and $\hat{G}_{QS}^{PR} = \frac{1}{8}S_8\!\left[\XX{P}{Q}\XX{R}{S}\right]$ terms must mutually commute. 

There is a correspondence between Pauli products and Majorana operators such that if two Majorana operators commute, so do the corresponding Pauli products  \cite{bettaque_structure_2024}. Thus, we examine the Majorana representation of the $\hat{G}_i$ operators to determine if mutual commutativity holds. We consider Majorana operators defined as \cite{landahl_logical_2023}:
\bea
&\gamma_{P0} = \AN{P} + \CR{P} ,
&\;\;\;\;\gamma_{P1} = -i(\AN{P} - \CR{P}) 
\eea
These majorana operators anticommute:
\bea
\{\gamma_a, \gamma_b\} = 2\delta_{ab},
\eea
where $a$ and $b$ are composite indices of $P$ and $0$ or $1$.

We first consider the Majorana terms within a $\hat{G}_{Q}^{P}$ operator. The excitation operators in the Majorana representation are:
\begin{align}
    \XX{P}{Q} &= \frac{1}{4}\left(\gamma_{P0}-i\gamma_{P1}\right)\left(\gamma_{Q0}+i\gamma_{Q1}\right)\\
    &=\frac{1}{4}\left(\gamma_{P0}\gamma_{Q0}+i\gamma_{P0}\gamma_{Q1}-i\gamma_{P1}\gamma_{Q0}+\gamma_{P1}\gamma_{Q1}\right) \\    
    \XX{Q}{P} &= \frac{1}{4}\left(\gamma_{Q0}-i\gamma_{Q1}\right)\left(\gamma_{P0}+i\gamma_{P1}\right)\\
    &=\frac{1}{4}\left(\gamma_{Q0}\gamma_{P0} +i\gamma_{Q0}\gamma_{P1} -i\gamma_{Q1}\gamma_{P0} +\gamma_{Q1}\gamma_{P1}\right)
\end{align}
Then, we have 
\begin{align}
    \hat{G}_{Q}^{P} &= \frac{1}{2}\left(\XX{P}{Q} + \XX{Q}{P}\right) \\
    &= \frac{1}{8}\left(\gamma_{P0}\gamma_{Q0}+i\gamma_{P0}\gamma_{Q1}-i\gamma_{P1}\gamma_{Q0}+\gamma_{P1}\gamma_{Q1}\right.\nonumber\\
    &\left.+\gamma_{Q0}\gamma_{P0} +i\gamma_{Q0}\gamma_{P1} -i\gamma_{Q1}\gamma_{P0} +\gamma_{Q1}\gamma_{P1}
    \right)\\
    &= \frac{1}{8}\left(
    i\gamma_{P0}\gamma_{Q1}-i\gamma_{Q1}\gamma_{P0}-i\gamma_{P1}\gamma_{Q0}+i\gamma_{Q0}\gamma_{P1}
    \right)\\
    &= \frac{i}{4}\left(
    \gamma_{P0}\gamma_{Q1}-\gamma_{P1}\gamma_{Q0}
    \right)
\end{align}
where $\left[
    \gamma_{P0}\gamma_{Q1},\gamma_{P1}\gamma_{Q0}
    \right]=0$ and we note the inclusion of both $\XX{P}{Q}$ and $\XX{Q}{P}$ are necessary to cancel out otherwise non-mutually commuting terms.

We then follow the same procedure with a $\hat{G}_{QS}^{PR}$ operator. Given the large number of terms involved, we use OpenFermion \cite{mcclean_openfermion_2019} to define Majorana operators and perform the algebraic manipulations. As discovered from these symbolic calculations, we need to include all eight terms that correspond to the eightfold permutation symmetery of the two body tensor $g_{PQRS}$ to cancel out several otherwise non-mutually-commuting Majorana terms. Given such cancellations, the corresponding Pauli products are guaranteed to mutually commute. We have placed the code in the notebook 
\url{licop_mutual_commutativity.ipynb} 
on the public GitHub repository \href{https://github.com/jtcantin/planted_solutions/tree/main/licop_mutual_commutativity}{jtcantin/planted\_solutions}.

We now know that $\hat{G}_{Q}^{P}$ and $\hat{G}_{QS}^{PR}$ \textit{individually} fulfill the mutual commutativity requirements of the underlying Pauli products. Furthermore, for any two operators that act on disjoint sets of orbitals, we know the mutual commutativity requirements of the underlying Pauli products are met as all of the terms contain an even number of Majorana operators. The question remains if there are terms sharing one or more indices, such as $\hat{G}_{QR}^{PR}$ and $\hat{G}_{UR}^{TR}$, which also fulfill the mutual commutativity requirements. While we did not check all of the possible patterns, we have shown (see the notebook mentioned above), that a pair of terms like $\hat{G}_{QR}^{PR}$ and $\hat{G}_{UR}^{TR}$, or even $\hat{G}_{QS}^{PQ}$ and $\hat{G}_{QT}^{QU}$, fulfill the requirements. This allows us to build a set of mutually-commuting $\hat{G}_i$ operators that span and couple a large number of orbitals.

\section{Hamiltonian Data Files and Computational Details}
\label{app:UUIDs}
The original Hamiltonian descriptions and data files can be found as part of the QB GSEE benchmark \cite{qb_benchmark_repo,gsee_benchmark_paper,qb_benchmark_repo}. The three bridged molybdenum Hamiltonians are documented in \url{problem_instance.mo2_n2.90d4e4fc-1216-4846-b45f-198c0530e29b.json} and correspond to the following UUIDs:
\begin{itemize}
    \item Task: e3a07092-d1e5-4867-a4d7-d0258f9df6db
    \newline FCIDUMP: 09c3ddd5\nobreakdash-\hspace{0pt}0187\nobreakdash-\hspace{0pt}46e8\nobreakdash-\hspace{0pt}95c3\nobreakdash-\hspace{0pt}157c470cb69a
    \item Task: 138733f0-08f5-4077-b848-813c8ec53c79
    \newline FCIDUMP: f36a9dbc\nobreakdash-\hspace{0pt}c34a\nobreakdash-\hspace{0pt}401f\nobreakdash-\hspace{0pt}8ea6\nobreakdash-\hspace{0pt}dc996d785edf
    \item Task: 79b74ad4-afa8-458f-87a3-a5ec339056c6
    \newline FCIDUMP: 9cba211b\nobreakdash-\hspace{0pt}820c\nobreakdash-\hspace{0pt}4ad4\nobreakdash-\hspace{0pt}a050\nobreakdash-\hspace{0pt}336e8049e1c7
\end{itemize}

The three Mn mono Hamiltonians are listed in \url{problem_instance.mn_mono.cb40f3f7-ffe8-40e8-4544-f26aad5a8bd8.json} and correspond to the following UUIDs:
\begin{itemize}
    \item Task: 6c39d3a7-c71a-4049-a8c7-3a3a4b61d8da 
    \newline FCIDUMP: 28a7820f-63fe-4920-aeec-a7ffe7e55d83
    \item Task: a16168e2-98cd-430d-a180-84f64b4d5e75
    \newline FCIDUMP: bae2da57\nobreakdash-\hspace{0pt}6a69\nobreakdash-\hspace{0pt}483e\nobreakdash-\hspace{0pt}95bc\nobreakdash-\hspace{0pt}b77f72ebfba8
    \item Task: 6e617348-d0b3-4b6d-9923-ea54ba5cf751
    \newline FCIDUMP: ea55abec\nobreakdash-\hspace{0pt}8253\nobreakdash-\hspace{0pt}445d\nobreakdash-\hspace{0pt}85fa\nobreakdash-\hspace{0pt}914948b5e5a5
\end{itemize}

The DMRG calculations are run on the Niagara cluster hosted by SciNet \cite{ponce2019deploying,loken2010scinet}.  SciNet is partnered
with Compute Ontario and the Digital Research Alliance of Canada. 188 GiB of RAM and 40
Intel “Skylake” 2.4 GHz cores or 40 Intel “CascadeLake” 2.5 GHz cores are available on each node of Niagara.

\section{Planted Solution Construction}
\label{app:construction}

As discussed in Sec.~\ref{sec:general_features}, there are at least three general ways of selecting the planted solution parameters that are applicable to all of the planted solution classes: (i) choose the parameters arbitrarily, (ii) extract the parameters from an extant Hamiltonian, and (iii) optimize over the parameters to fit the planted solution to an extant Hamiltonian. This appendix expands on some of the methods for specific planted solution classes. 

\subsection{MF}
For all of the construction methods of the MF planted solutions (Sec.~\ref{sec:HFsol}), we first combine the one-electron term with the two-electron term in the spin-orbital basis (see  Appendix~\ref{app:one_el}).

Here, we describe an extraction construction method based on the double factorization (DF) decomposition of a Hamiltonian. 
With a Cholesky decomposition of the two-electron tensor, the electronic
Hamiltonian can be expressed as
\begin{equation} \label{eq:H_DF}
    \hat H_\mathrm{DF} = \sum_{m=1}^M c_m U_m^\dagger \left( \sum_P \epsilon_P^{(m)} \hat n_P \right)^2 U_m,
\end{equation}
with $M$ the total number of fragments, $c_m$ the coefficient for fragment $m$, and we order $|c_1| \geq |c_2| \geq ... \geq |c_M|$.
We choose the largest fragment to form the planted solution:
\begin{equation} \label{eq:H_DFLF}
    \hat H_{\rm{DFLF}} = U_1^\dagger \left( \sum_P \epsilon^{(1)}_P\hat n_P\right)^2 U_1,
\end{equation}
where DFLF is an abbreviation of  Double Factorization, Largest Fragment.
The construction of this planted solution only requires linear algebra routines for the Cholesky decomposition, making it extremely efficient. However, the corresponding $\lambda_{PQ}$ coefficients of Eq.~\eqref{eq:HFs} are rank-deficient, lowering the flexibility of the planted solution.

To improve upon this rank deficient approach, we incorporate a portion of the operators that originate from terms with $m>1$.
To do this, we convert all of the terms with $m>1$ to the same orbital frame as that of the first fragment. We then collect all the components of these terms that are diagonal in the $U_1$ frame to form the $\lambda_{PQ}$ coefficients. 

Transforming Eq.~\ref{eq:H_DF} into the $U_1$ frame, we have
\begin{align}
    U_1 \hat H_{\rm{DF}} U_1^\dagger &= c_1 \left(\sum_{P}\epsilon^{(1)}_P\hat n_P\right)^2 \nonumber \\
    &\ \ + \sum_{m=2}^M c_m \tilde U_m^\dagger \left(\sum_{P}\epsilon^{(m)}_P\hat n_P\right)^2 \tilde U_m, \label{eq:rotated_DF}
\end{align}
where we have defined $\tilde U_m \equiv U_m U_1^\dagger$. Considering how a given orbital rotation acts on number operators $U^\dagger\hat n_P U = \sum_{QR} U_{QP} U_{RP}^* \XX{Q}{R}$, Eq.~\eqref{eq:rotated_DF} becomes
\begin{align}\label{eq:DFp}
    U_1 \hat H_{\rm{DF}} U_1^\dagger &= c_1 \left(\sum_{P}\epsilon^{(1)}_P\hat n_P\right)^2 \nonumber \\
    &\ \ + \sum_{m=2}^M c_m \sum_{PQ} \epsilon^{(m)}_P\epsilon^{(m)}_Q \sum_{IJKL} a_{IJKL}^{(PQm)} \XX{I}{J} \XX{K}{L},
\end{align}
where we have defined $a_{IJKL}^{(PQm)} \equiv \tilde U^{(m)}_{IP} \tilde U^{(m)*}_{JP} \tilde U^{(m)}_{KQ} \tilde U^{(m)*}_{LQ}$. All the diagonal terms corresponding to $\XX{I}{I} \XX{K}{K}$ are then collected in a single fragment to form the ``boosted DF'' (BDF) planted solution:
\begin{equation}
\label{eq:boosted_DF}
    \hat H_{\rm{BDF}} = U_1^\dagger \left( \sum_{PQ} \lambda^{\rm{BDF}}_{PQ} \hat n_P\hat n_Q \right) U_1,
\end{equation}
where we have defined
\begin{equation}
    \lambda^{\rm{BDF}}_{PQ} \equiv c_1\epsilon_P^{(1)}\epsilon_Q^{(1)} + \sum_{m=2}^{M} c_m \sum_{RS}\epsilon^{(m)}_R \epsilon^{(m)}_S  |\tilde U^{(m)}_{PR}|^2 |\tilde U^{(m)}_{QS}|^2.
\end{equation}

More generalized notions of mean field Hamiltonians, which have Slater determinant eigenstates constructed using multiple orbital bases \cite{izmaylovHowDefineQuantum2021,patelExtensionExactlySolvableHamiltonians2024,patelQuantumMeasurementQuantum2025}, can also be considered for mean field solvable planted solutions.

\subsection{LICOP}
The aspect unique to the LICOP planted solution class (Sec.~\ref{sec:licop_hamiltonians}) is the choice of mutually-commuting operators. Ideally, this set should be as large as possible for maximum flexibility. 

We have experimentally seen that a sorted-insertion approach \cite{Crawford:2019tg} to finding the largest group of commuting operators of a given Hamiltonian results in the set of $\hat{G}_{P}^{P}$ and $\hat{G}_{PR}^{PR}$ operators receiving the largest 1-norm weight. Thus, a reasonable approach 
is to begin with the non-excitation operators $\hat{G}_{P}^{P}$ and $\hat{G}_{PR}^{PR}$.  
We then 
modify this set with mutation combined with importance selection.

Mutation involves modifying the operator to consist of one or more excitations. For example, $\hat{G}_{0}^{0}\rightarrow\hat{G}_{0}^{1}$ or $\hat{G}_{01}^{01}\rightarrow\hat{G}_{01}^{02}$. After mutation, we remove any non-excitation operators that no longer commute with the new operator or discard this new operator if it does not commute with previously-mutated operators. In general, this commutativity check scales as $O(N^2)$, but can be greatly sped up by recognizing that any two operators that act on disjoint sets of orbitals are guaranteed to commute. 

The probability distribution used for the importance selection will differ based on the construction method used. If the coefficients are chosen randomly, the distribution can be arbitrary. If we are taking the coefficients from an extant Hamiltonian, a good approach is to prioritize the operators based on the magnitude of their coefficients in the source Hamiltonian. If we are trying to fit the coefficients to a source Hamiltonian, the same selection approach would be beneficial.

\subsection{CASS}
As the CASS planted solutions (Sec.~\ref{sec:CASsol}) are composed of independent subsystems, they can be chosen in a variety of ways; so long as they are individually solvable within a reasonable time frame, that is. The independent subsystems could be the Hamiltonians of small molecules or of other planted solution types, for example. The primary approach we take is to choose a source Hamiltonian, choose a partitioning of the orbitals, and then eliminate all terms of the source Hamiltonian that would otherwise couple orbitals assigned to different subsystems. 

\subsection{AGP}
For the AGP Hamiltonians (Sec~\ref{sec:agp_hamiltonian}), we define $c_p\in \mathbb{R}$ and $d_{pq}^{(u)}\in\mathbb{R}$ to keep the Hamiltonians real. However, the construction also works with $c_p\in \mathbb{C}$ and $d_{pq}^{(u)}\in\mathbb{C}$.

\bibliographystyle{achemso}
\bibliography{library}

\end{document}